\documentclass[12pt]{article}
\pdfoutput=1
\usepackage[a4paper]{geometry}
\usepackage{jheppub, amsmath,amssymb,amsfonts,amsxtra, mathrsfs, makeidx,graphics,graphicx,amsthm,epsfig, ytableau,bm,longtable,float, color,tikz,mathtools,xfrac,footnote,rotating, lscape, makecell, environ,mathtools, empheq, colortbl}
\usepackage{slashed}

\usetikzlibrary{positioning}
\usetikzlibrary{chains}
\usetikzlibrary{arrows,fit,decorations.pathreplacing}
\tikzstyle{every picture}+=[remember picture]
\tikzstyle{na} = [baseline]

\usetikzlibrary{arrows, decorations.markings, calc, fadings, decorations.pathreplacing, patterns, decorations.pathmorphing, positioning}

\tikzstyle{every picture}+=[remember picture]
\tikzstyle{na} = [baseline=-.5ex]

\usepackage{bbm}
\usepackage{subfigure}

\numberwithin{equation}{section}

\newcommand{\be}{\begin{equation}} \newcommand{\ee}{\end{equation}}
\newcommand{\bea}{\begin{equation} \begin{aligned}} \newcommand{\eea}{\end{aligned} \end{equation}}

\def\tilde{\widetilde}

\def\bar{\overline}
\def\a{\alpha}

\def\rt2{\sqrt{2}}

\def\Re{\mathop{\rm Re}}
\def\Im{\mathop{\rm Im}}

\def\rmd{{\rm d}}

\def\pd{\partial}

\def\1{{\ds 1}}

\def\repa{\raise4pt\hbox{$\square$}\mkern-14mu\raise-4pt\hbox{$\square$}}
\def\repab{\overline{\raise4pt\hbox{$\square$}\mkern-14mu\raise-4pt\hbox{$\square$}\mkern-1mu}}

\def\smileface{\ensuremath{\hbox{\large$\bigcirc$}\mkern-15mu\raise-1pt\hbox{\scriptsize$\smallsmile$}%
\mkern-10mu\raise4pt\hbox{..}\mkern4mu}}
\def\frownface{\ensuremath{\hbox{\large$\bigcirc$}\mkern-15mu\raise-1pt\hbox{\scriptsize$\smallfrown$}%
\mkern-10mu\raise4pt\hbox{..}\mkern4mu}}

\newcommand{\ba}{\begin{array}}
\newcommand{\ea}{\end{array}}
\newcommand{\bi}{\begin{itemize}}
\newcommand{\ei}{\end{itemize}}

\def\bea#1\eea{\allowdisplaybreaks \begin{align}#1\end{align}}
 \newcommand{\ben}{\begin{enumerate}}
\newcommand{\een}{\end{enumerate}}
\newcommand{\bean}{\begin{eqnarray*}}
\newcommand{\eean}{\end{eqnarray*}}

\newcommand{\comment}[1]{}

\definecolor{light-gray}{gray}{0.7}

\def\aup#1 {\overset{#1}{\uparrow} \, \overset{\tilde{#1}}{\downarrow}}

\tikzset{snake it/.style={decorate, decoration={snake, amplitude=.4mm, segment length=2mm,
                       post length=0mm,pre length=0mm}}}

\newcommand{\Dt}{\mathrm{D3}}
\newcommand{\aDt}{\overline{\mathrm{D3}}}
\newcommand{\Dse}{\mathrm{D7}}

\newcommand{\UV}{\mathrm{UV}}
\newcommand{\IR}{\mathrm{IR}}
\newcommand{\KS}{\mathrm{KS}}

\def\YM{\mathrm{Y}\mathrm{M}}

\title{$\overline{\text{D3}}$-branes and gaugino condensation
}
\author[a]{Iosif Bena,}
\author[b]{Emilian Duda\c s,} 
\author[a]{Mariana Gra\~na,}
\author[a]{Gabriele Lo Monaco}
\author[a]{and Dimitrios Toulikas}
\affiliation[a]{Institut de Physique Th\'eorique, Universit\'e Paris Saclay, CEA, CNRS, \\ Orme des Merisiers, 91191 Gif-sur-Yvette CEDEX, France}
\affiliation[b]{CPHT, CNRS, Ecole Polytechnique, IP Paris, \\ F-91128 Palaiseau, France}
\emailAdd{iosif.bena@ipht.fr}
\emailAdd{mariana.grana@ipht.fr}
\emailAdd{gabriele.lomonaco@ipht.fr}
\emailAdd{emilian.dudas@polytechnique.edu}
\emailAdd{dimitrios.toulikas@ipht.fr}

\abstract{Anti-D3 branes at the bottom of warped throats, commonly used to uplift the cosmological constant in String-Theory de Sitter proposals, source a pl\ae thora of supersymmetry-breaking fluxes, that can interact nontrivially with other ingredients of the flux compactification. In this paper we perform a complex-structure decomposition of these fluxes, and compute the effect of the (0,3) flux component on the stabilization of K\"ahler moduli via D7-branes gaugino condensation. This allows us to obtain a new constraint on the validity of this stabilization mechanism. This effect does not appear hard to satisfy in de Sitter construction proposals that use long warped throats, but may be problematic in proposals where the warping is small.}

\begin{document}
\maketitle


\section{Introduction}

Many proposals to construct de Sitter vacua in String Theory involve uplifting the negative cosmological constant that one typically obtains in flux compactifications after fixing the moduli. The original and most popular uplifting ingredient are anti-D3 branes placed in long warped throats within the compactification manifold \cite{Kachru:2003aw}. They are argued to give a tunably small uplift energy, which would make the cosmological constant positive without perturbing the stabilization of the other moduli. 

The prototypical example of such a warped throat is the Klebanov-Strassler (KS) geometry, obtained by adding fluxes to the deformed conifold \cite{Klebanov:2000hb}. There is by now an extensive body of work investigating the physics of anti-D3 branes in the KS geometry, in several regimes of parameters. Many of the results of these calculation point towards the existence of pathologies and instabilities when multiple anti-D3 branes are placed at the bottom of the KS geometry \cite{Bena:2011wh,Bena:2009xk,Bena:2014jaa,Danielsson:2014yga,Blumenhagen:2019qcg}. Furthermore, if the flux on the three-cycle of the deformed conifold is not large, even a single anti-D3 brane appears to have the power to cause a runaway behavior in the conifold deformation modulus  \cite{Bena:2018fqc,Blumenhagen:2019qcg,Randall:2019ent,Dudas:2019pls}, collapsing the whole KS geometry into the singular Klebanov-Tseytlin one \cite{Klebanov:2000nc} and annihilating against the singularity.\footnote{For a recent argument against this scenario see \cite{Lust:2022xoq}.}  A similar behavior is found when considering black holes in the KS geometry\cite{Buchel:2018bzp,Bena:2019sxm}. Hence, the only corner of parameter space where KS antibranes have any chance of being metastable is when there is a single antibrane and the flux on the three-cycle of the deformed conifold is large.

However, even a single anti-brane has a significant back-reaction, and is capable of significantly affecting the other ingredients of the de Sitter construction. It is the purpose of this paper  to calculate the strength the interactions between this single antibrane at the bottom of a KS solution and other ingredients that are used in the KKLT proposal \cite{Kachru:2003aw} to construct de Sitter vacua with stabilized moduli in String Theory. 

If we start from a supersymmetric solution in which a KS throat is glued to a  CY geometry, the complex structure of the KS throat will match the complex structure of the bulk compact CY.  Hence, the fluxes sourced by the anti-D3 branes can be decomposed according to the complex-structure of the KS throat and, as expected, have all possible components. The values of the field strength fluxes at the top of the KS geometry, where it is glued to the compactification manifold, are expected to be of the same order as the values of these fluxes in the rest of this manifold, away from the throat, and they determine the magnitude of the effect of  antibranes on the other ingredients needed in de Sitter construction proposals. 

As we have shown in a previous paper \cite{Bena:2022cwb}, the $(0,3)$ component of the fluxes sourced by the anti-D3 branes gives rise to a nontrivial constant term in the superpotential. This avoids  the need of one of the ingredients of the KKLT proposal, namely a finite but very small constant term in the superpotential, introduced ad hoc by turning on (0,3) fluxes. Hence, taking  the interactions between antibranes and the rest of the flux-compactification ingredients fully into account can result in simplified ``bare-bones'' de Sitter proposals \cite{Bena:2022cwb}, which use one less ingredient than KKLT. 

Another possible consequence of the antibrane fluxes is to affect gaugino condensation on D7 branes wrapping certain divisors of the CY geometry. In the absence of D7 branes, the volumes of these divisors (which correspond to K\"ahler moduli of the compactification) are flat directions. The low-energy physics of the D7 branes wrapping a holomorphic divisor is an ${\cal N}=1$ Super-Yang-Mills theory, which confines in the infrared. The non-perturbative Affleck-Dine-Seiberg superpotential \cite{Affleck:1983mk} of this confining theory depends nontrivially on the K\"ahler moduli, and gives rise to a term in the potential that is responsible for stabilizing the size of the divisors \cite{Kachru:2003aw}.

The fields sourced by the anti-D3 brane can in principle interfere with this sensitive mechanism of K\"ahler-moduli stabilization. Indeed, as shown in \cite{Camara:2004jj,Jockers:2004yj,Grana:2020hyu}, both the (1,2) and the (0,3) components of the complex three-form field strength give rise to mass terms for the fermions on the D7 brane. These mass terms break the supersymmetry of the ${\cal N}=1$ theory on the D7 branes, and can modify the RG flow and the gaugino-condensation scale. More precisely, they can affect the Affleck-Dine-Seiberg superpotential and can introduce extra terms in the potential for the K\"ahler moduli, potentially ruining their stabilization. 

In this paper we use the expression of the supersymmetry-breaking fluxes sourced by the anti-D3 branes \cite{Bena:2011hz,Bena:2011wh} to calculate precisely their effect on the stabilization of K\"ahler moduli via D7-brane gaugino condensation. We obtain a new bound relating various parameters of compactification with long warped throats, which must be satisfied in order to be able to stabilize K\"ahler moduli via D7-brane gaugino condensation. We  combine this bound with other constraints appearing in vanilla-type KKLT compactifications, and find that it is generically satisfied. We leave the exploration of the importance of this bound on de Sitter proposals that do involve large warping\footnote{Such as the proposals in \cite{Crino:2020qwk,Bento:2021nbb}} to future exploration.

Beside their effect on the gaugino condensation on D7 branes, the extra fluxes and fields sourced by the antibranes can also affect the number of fermion zero modes on the D3 instantons that also give rise to non-perturbative terms in the superpotential \cite{Witten:1996bn} that can stabilize other K\"ahler moduli, possibly switching off these terms. This is a subtle effect \cite{Bergshoeff:2005yp,Lust:2005cu}\footnote{See also \cite{Robbins:2004hx,Tripathy:2005hv,Kallosh:2005gs, Saulina:2005ve,Martucci:2005rb, Berglund:2005dm}.}, which potentially can be more drastic than the effect on D7 gaugino condensation: even the tiniest ``wrong fluxes'' sourced by the antibranes are enough to uplift fermion zero modes and ruin the stabilization of certain K\"ahler moduli. We leave its investigation for future work.

As a byproduct of our calculation, we decode some aspects of the holographic dictionary corresponding to anti-D3 branes in the Klebanov-Strassler geometry. If a single anti-D3 brane in the KS geometry with large three-form flux on the $S^3$ of the deformed conifold is indeed metastable, the resulting solution would be holographically dual to a metastable vacuum of the quiver gauge theory dual to the KS geometry. The structure of vacua of this theory is quite rich \cite{Dymarsky:2005xt}, and the existence of a metastable vacuum at strong coupling would be a nontrivial prediction of holography. It would be very interesting to try construct this putative vacuum directly in field theory, using for example ISS methods \cite{Intriligator:2006dd}. The fact that this vacuum can only exist in a very restricted region of parameter space may also explain why earlier attempts at finding it have not been successful \cite{Seiberg-private-communication}.

Our analysis gives several clear holographic indications as to what the physics of this vacuum is. In particular, the fall-offs of the three-form fluxes with the radius can be used to show that certain dimension-three operators corresponding to fermion bilinears and certain dimension-seven operators corresponding to fermion bilinears multiplied by $F_{\mu \nu} F^{\mu \nu}$ acquire non-trivial vacuum expectation values. Furthermore, the $(1,2)$ fluxes sourced by the antibranes fall off as $1/r^4$, and hence give rise to a nontrivial vacuum expectation value of a dimension-4 operator corresponding to a marginal deformation of the superpotential \cite{Grana:2001xn}. Since the dictionary between the bulk three-form fields and the fermion bilinears is well understood \cite{Kuperstein:2003yt, Aharony:2005zr}, we believe this information will be useful in searching for the holographic dual of the putative KS metastable vacuum.

The paper is organized as follows: In Section \ref{sec:KSgeometry} we review the Klebanov-Strassler geometry, as well as the most general deformation preserving its $SU(2) \times SU(2) \times \mathbb{Z}_2$ symmetry and describe how the solution corresponding to anti-D3 branes smeared at the tip of the throat is obtained. In Section \ref{sec:fluxzoo} we write the analytic expression, as well as the UV expansion, of all $G_3$ flux components that are generated by the addition of the anti-D3 branes and comment on their holographic duals. In Section \ref{sec:gauginomass} we compute the D7-brane gaugino mass induced by the $G_{(0,3)}$ flux sourced by the anti-D3 branes and compare it with a four-dimensional supergravity description of supersymmetry-breaking gaugino masses finding parametric agreement. Moreover, we derive a bound that all KKLT-like constructions should satisfy in order for the K\"ahler moduli stabilization via gaugino condensation to work, and argue that it is easily satisfied by the existing constructions.


\section{$\overline{\text{D3}}$formed KS geometry}
\label{sec:KSgeometry}
\subsection{Review of the $\KS$ solution and its non-supersymmetric deformations}
A long warped Klebanov-Strassler-like throat, at the bottom of which anti-$\Dt$ branes (denoted as $\aDt$-branes in the following) can sit, is a key element in the KKLT proposal for constructing de Sitter vacua in String Theory \cite{Kachru:2003aw}. In this section we review the supersymmetric Klebanov-Strassler (KS) geometry \cite{Klebanov:2000hb}, as well as the ansatz that describes the most general deformation (with vanishing RR axion $C_0$) that preserves its $SU(2)\times SU(2)\times \mathbb{Z}_2$ symmetry \cite{Papadopoulos:2000gj}. The ten-dimensional spacetime consists of a warped product of four-dimensional Minkowski space and the deformed conifold :
\be
\label{eq:metricKS}
\text{d}s^2_{10}\,=\,e^{2A+2W-X}\text{d}s_{1,3}^2 \,+\,e^{-6W-X}\text{d}\tau^2\,+\,e^{X+Y}(g_1^2+g_2^2)\,+\,e^{X-Y}(g_3^2+g_4^2)\,+\,e^{-6W-X}g_5^2\,,
\ee
where $\{A,W, X,Y\}$ are functions of the radial coordinate, $\tau$, and the one-forms $g_i$ are: 
\be
\begin{split}
&g_1\,=\,-\frac{1}{\sqrt2}\Im(w_1+w_2)\,,\quad  g_2\,=\,\frac{1}{\sqrt 2}\Re(w_1-w_2)\,,\\
&g_3\,=\,-\frac{1}{\sqrt2}\Im(w_1-w_2)\,,\quad  g_4\,=\,\frac{1}{\sqrt 2}\Re(w_1+w_2)\,,\\
& g_5\,=\,\rmd\psi+\sum_{i=1}^2 \cos \theta_i \rmd\phi_i\,, \quad  g_6\,=\,\rmd\tau\,,
\end{split}
\ee
with $0\leq \psi \leq 4\pi,\, 0\leq \theta_i \leq \pi,\, 0\leq \phi_i \leq 2\pi$. Here we introduced the forms $w_1\,\equiv \,\rmd \theta_1+i \sin \theta_1\,\rmd \phi_1$ and $e^{i\psi}w_2\,\equiv \,\rmd \theta_2+i \sin \theta_2\,\rmd \phi_2$, with $\mathbb{Z}_2$ exchanging the two $S^2$'s defined by $w_i$ \cite{Papadopoulos:2000gj}.

The NSNS and RR forms, $H_3$, $F_3$ and $F_5$, are all non-vanishing and their form is fixed by Bianchi identities and isometries:
\be
\begin{split}
\label{eq:fluxesKS}
H_3\,&=\,\frac{1}{2}(k-f)g_5\wedge(g_1\wedge g_3+g_2\wedge g_4)\,+\,\text{d}\tau\wedge\left(f'\,g_1\wedge g_2\,+\,k'\,g_3\wedge g_4 \right)\,,\\
F_3\,&=\,F\,g_1\wedge g_2\wedge g_5\,+\,(2P-F)\,g_3\wedge g_4\wedge g_5\,+\,F'\,\text{d}\tau\wedge(g_1\wedge g_3+g_2\wedge g_4)\,,\\
F_5\,&=\,\left[ \frac{\pi\,Q}{4}+(k-f)F+2P\,f \right](1+\star_{10})\,g_1\wedge g_2\wedge g_3\wedge g_4\wedge g_5\,, \\
\Phi\, &=\,\Phi(\tau)\,, \quad C_0 \,=\,0
\end{split}
\ee
with $Q$ a constant and $\{k,f, F\}$ functions of $\tau$. 

The radial dependence of the functions appearing in the KS metric is:
\be
\begin{split}
\label{eq:funcKS}
e^{X_{KS}}\,&=\,\frac{1}{4}h(\tau)^{1/2}\left(\frac{1}{2}\sinh(2\tau)-\tau\right)^{1/3}\quad\,\,\,\,\,\,\,,\quad e^{6A_{KS}}\,=\,S^2\frac{\sinh(\tau)^2}{3\cdot 2^5}e^{2X_{KS}}\,,\\
e^{6W_{KS}}\,&=\,\frac{24}{h(\tau)\sinh(\tau)^2}\left(\frac{1}{2}\sinh(2\tau)-\tau\right)^{1/3}\,,\quad e^{Y_{KS}}\,=\,\tanh(\tau/2)\,,
\end{split}
\ee
where $S$ is a complex-structure modulus and $h(\tau)$ is the solution that vanishes at infinity to the following differential equation:
\be
\frac{\text{d}h}{\text{d}\tau}\,=\,32\,P^2\,g_s\,\frac{\tau \coth\tau-1}{\sinh(\tau)^2}\left(\frac{1}{2}\sinh(2\tau)-\tau\right)^{1/3}\,,
\ee
with $P=\frac{1}{4}\a' M$ and $M$ the number of units of $F_3$ flux on the large $\,\mathrm{S}^3$ of the warped deformed conifold (also known as the compact $A$-cycle). To simplify the expressions, it is useful to define the constants $h_0\equiv h(\tau\!=\!0)$ and $\mathcal{I}_0\equiv\frac{h_0}{32P^2 g_s}$. 

Moreover, the functions $\{f,k,F\}$ appearing in the fluxes are given by:
\be
\begin{split}
f_{KS}\,&=\,-g_s\,P\,\frac{(\tau\coth(\tau)-1)(\cosh(\tau)-1)}{\sinh(\tau)}\,,\\
k_{KS}\,&=\,-g_s\,P\,\frac{(\tau\coth(\tau)-1)(\cosh(\tau)+1)}{\sinh(\tau)}\,,\\
F_{KS}\,&=\,\,P\,\frac{\sinh(\tau)-\tau}{\sinh(\tau)}\,,
\end{split}
\ee
and, when there are not mobile D3 branes, $Q$ is zero.

It is worth noting that in the KS solution the complexified three-form, $G_3$, satisfies the ISD condition, $(i+\star_6)G_3=0$. Furthermore, $G_3$ has only  $(2,1)$ components with respect to the choice of holomorphic vielbeins picked by supersymmetry :
\be
\label{eq:complexStructure}
\begin{split}
h_1\,&=\,E_1+i\,(\cos \omega E_2+\sin\omega\,E_4)\,,\\
h_2\,&=\,E_3+i\,(\sin \omega E_2-\cos\omega\,E_4)\,,\\
h_3\,&=\,E_5+i\,E_6\,,
\end{split}
\ee
where $\sin{\omega}=-\tanh Y$ and:
\be
\begin{split}
E_{1}\,&=\, \frac{e^{X/2}}{\sqrt{2\cosh(Y)}}(g_2+g_4)\,,\quad E_{2}\,=\, \frac{e^{X/2}}{\sqrt{2\cosh(Y)}}(g_1+g_3)\,,\\
E_{3}\,&=\, \frac{e^{X/2}\sqrt{\cosh(Y)}}{\sqrt{2}}(g_4-g_2-(g_2+g_4)\tanh{Y})\,,\\
E_{4}\,&=\, \frac{e^{X/2}\sqrt{\cosh(Y)}}{\sqrt{2}}(g_3-g_1-(g_1+g_3)\tanh{Y})\,,\\
E_{5}\,&=\, e^{-X/2-3W}\text{d}\tau\,,\quad E_{6}\,=\, e^{-X/2-3W}g_5\,.
\end{split}
\ee
We will denote the holomorphic 3-form on the {\em warped} geometry as $\Omega \equiv n \, h_1\wedge h_2\wedge h_3$, where the normalization constant, $n$, is such that $\Omega^{\rm KS}$, the unwarped Klebanov-Strassler 3-form, satisfies:
\be
\label{eq:OmegaNorm}
\int_A H^{-3/4}\Omega^{KS}\!=\!S \quad \Rightarrow \quad n\!=\!-\frac{\sqrt{6}}{4\pi^2}\,,
\ee
where $H=e^{-4A-4W+2x}$ is the warp factor\footnote{The KS metric \eqref{eq:metricKS} can be written in terms of $H$ as $ds^2=H^{-1/2}ds^2_4+H^{1/2}ds^2_6$.}. The explicit form of the normalized $\Omega^{KS}$ in terms of the set of  $\{g_i\}$ is the following: 
\be
\label{eq:OmegaKS}
 \Omega^{\KS}\!=\!-\frac{h(\tau)^{3/4}\sinh\tau}{16\pi^2}\!\left( \tanh\!\left(\tfrac{\tau}{2}\right)\!g_1\!\wedge\! g_2\!-\!\coth\!\left(\tfrac{\tau}{2}\right)\! g_3\!\wedge \!g_4\!+\!i g_1\!\wedge \!g_3+\!i g_2\! \wedge\! g_4\right)\wedge(g_5-i g_6)\,.
\ee
Note that with this normalization constant, $\Omega \wedge \bar{\Omega}=\frac{3i}{\pi^4}\text{vol}_6$, instead of $\Omega \wedge \bar{\Omega}=8i\text{vol}_6$, which is the usual convention in the literature\footnote{Remember that $\text{vol}_6$ is the volume of the six-dimensional internal space including the warp factor.}.
\subsection{Adding $\aDt$-branes}
Anti-$\Dt$ branes have a charge that is opposite to that of the Klebanov-Strassler geometry and, in the probe approximation, fall to the $S^3$ tip of the throat. When the antibranes are localized on the $S^3$, the fields they source have a complicated dependence on $\tau$ and the $S^3$ coordinates, and are hard to compute analytically. However, if we are interested in the solution away from the tip, one can assume the antibranes to be smeared, and then the solution will have $SU(2)\times SU(2)\times \mathbb{Z}_2$ symmetry. Its metric and fluxes will then be described by the eight functions of $\tau$ appearing in the Papadopoulos-Tseytlin ansatz \eqref{eq:metricKS}-\eqref{eq:fluxesKS}:
\be
\{\Phi^i\}\,=\,\{X-2W-5A,\,Y,\,X+3W,\,X-2W-2A,\,f,\,k,\,F,\,\Phi\}\,,
\ee
where we performed the above redefinition for convenience. 

When the number of the anti-D3 branes, $\overline{N}$,  is small one can describe their solution as a small perturbation around the KS geometry \cite{Bena:2011hz,Bena:2011wh,Dymarsky:2011pm,Dymarsky:2013tna}:
\be
\Phi^i\,=\,\Phi^i_{\KS}\,+\, \varepsilon \,\varphi^i\,+\,O(\varepsilon^2)
\ee
where $\varepsilon$ is an expansion parameter that can be taken to be:
\be
\varepsilon=\frac{\overline{N}}{g_sM^2}\,.
\ee

We then require the equations of motion of type-IIB supergravity to be satisfied at leading order in $\varepsilon$. Plugging the particular ansatz \eqref{eq:metricKS}-\eqref{eq:fluxesKS} in the type IIB supergravity action, one gets an action for the fields $\{\Phi^i\}$ that can be cast in the following form \cite{Papadopoulos:2000gj}:
\be
\mathcal{L}\,=\,-\frac{1}{2}\mathrm{G}_{ij}\left(\frac{\rmd\Phi^i}{\rmd\tau}-\frac{1}{2}\mathrm G^{i k}\frac{\pd W}{\pd\Phi^k}\right)\left(\frac{\rmd\Phi^j}{\rmd\tau}-\frac{1}{2}\mathrm G^{j l}\frac{\pd W}{\pd\Phi^l}\right)-\frac{1}{2}\frac{\pd W}{\pd\tau}\,,
\ee
where $G$ and $W$ are respectively a $\Phi$-dependent metric and superpotential whose exact functional dependence is not relevant in the following. To study perturbations  around a supersymmetric solution governed by this action \cite{Borokhov:2002fm} it is useful to introduce the set of functions, $\{\xi^i\}$, conjugate to the perturbations, $\{\varphi^i\}$. They are defined as:
\be
\xi_i\,\equiv \,\mathrm{G}_{ij}(\Phi_{\KS})\left(\frac{\rmd \varphi^j}{\rmd \tau}-\mathrm{M}^{j}{}_k\left(\Phi_\KS\right)\varphi^k\right)\,,\quad \mathrm{M}^{j}{}_k\,\equiv\,\frac{1}{2}\frac{\pd}{\pd\Phi^k}\left(\mathrm{G}^{jl}\frac{\pd W}{\pd\Phi^l}\right)\,.
\ee
The supersymmetric $\KS$ background corresponds to $\xi_i=0$ for all $i$. Our goal is to find other solutions to the equations of motion:\footnote{These equations are actually supplemented by the zero-energy condition $\xi_i \frac{\rmd \Phi^i_\KS}{\rmd \tau}=0$.}
\be
\begin{split}
\frac{\rmd\xi_i}{\rmd \tau}\,+\,\xi_j \mathrm{M}^{j}{}_i(\Phi_{\KS})\,&=\,0\,,\\
\frac{\rmd\varphi^i}{\rmd \tau}-\mathrm{M}^i{}_j(\Phi_{\KS})\varphi^{j}\,&=\,\mathrm{G}^{ij}\xi_j\,.
\end{split}
\ee

An analytical form for the most general perturbation can be found and it involves various nested integrals \cite{Bena:2011hz}. Nevertheless, such integrals can be evaluated both as series expansions in the UV or IR limits, or numerically throughout the whole solution. The numerical evaluation allows one to match the parameters of the UV and IR expansions. The simplest example of this matching is the evaluation of the ``momentum" $\xi_1$, which controls the force on a probe D3-brane:
\be
F_{\mathrm{D}3}=\frac{2\varepsilon}{3}e^{-2X_{\KS}}\xi_1\,,\quad \xi_1\,=\,X_1\,h(\tau)\,,
\ee
where $X_1=\frac{3\pi}{8\mathcal{I}_0}\frac{S^{4/3}}{h_0}$ is an integration constant. As we have already observed, the function $h(\tau)$ can be expressed as a definite integral, and its value at an arbitrary $\tau$ can only be evaluated numerically. However, its asymptotic IR and UV expansions can be evaluated straightforwardly:
\be
h_{\IR}=h_0-\frac{16}{3}\left(\frac{2}{3}\right)^{\frac{1}{3}}g_s P^2\,\tau^2\,+\,O(\tau^4)\,,\quad \!\!\!\! h_{\UV}=12\times 2^{\frac{1}{3}}g_s P^2(4\tau-1)e^{-4\tau/3}\,+\,O(e^{-10\tau/3}).
\ee
The expressions for the other perturbations and momenta are considerably more involved and can be found in \cite{Bena:2011wh}.

\section{The flux zoo}
\label{sec:fluxzoo}
When the KS throat is deformed because of the presence of anti-D3 branes, the $G_3$ flux takes the most general form:
\be
G_3\,=\, G_{(3,0)}\,+\,G_{(0,3)}\,+\, G_{(1,2)}\,+\, G_{(2,1)}\,.
\ee
The first three components, which break supersymmetry, are sourced by the anti-D3 branes and appear at $O(\varepsilon)$. This happens because the fluxes $H_3$ and $F_3$, defined  in \eqref{eq:fluxesKS}, do not combine anymore into a $(2,1)$ form (with respect to the complex structure \eqref{eq:complexStructure} of the zeroth order KS solution) for a generic choice of the functions $\Phi^i$.

Using the complex structure \eqref{eq:complexStructure}, we can extract the various components of $G_3$. For example, the $(0,3)$ component of $G_3$ for an arbitrary set of $\Phi^i$ is:
\be
\label{eq:gamma03a}
\begin{split}
G_{(0,3)}\,&=\,-\frac{1}{8}\,\gamma_{(0,3)}\,\overline{h}_1\wedge \overline{h}_2\wedge \overline{h}_3\,,\\
\gamma_{(0,3)}\,&=\,e^{3W-\phi-X/2}\left(2e^{\phi}\left(e^{Y}P-\cosh Y\,F-F'\right)+e^Y k'-e^{-Y}f'+k-f\right)\,.
\end{split}
\ee
This expression holds at any order in $\varepsilon$. The function $\gamma_{(0,3)}$ vanishes in the KS limit, while its first-order term in the $\varepsilon$ expansion is:
\be \label{eq:gamma03b}
\gamma_{(0,3)}=-\frac{8\sqrt6\,g_s^{-1}}{ h^{\tfrac{3}{4}}\sinh(\tau)^2}\pd_\tau\!\left(g_s \sinh\tau\,\varphi_7+\cosh(\tfrac{\tau}{2})^2\varphi_5-\sinh(\tfrac{\tau}{2})^2\varphi_6\right)+O(\varepsilon^2)\,.
\ee
Let us stress that in \eqref{eq:gamma03a} the vielbeins, $\overline{h}_i$, can be taken to be the Klebanov-Strassler ones, since any correction to them would come at order $O(\varepsilon^2)$. Thus, the first correction to $G_{(0,3)}$ is:
\be
\label{eq:G03}
G_{(0,3)}=\frac{\pi^2}{2\sqrt{6}}\gamma_{(0,3)}\overline{\Omega}^{\,\KS}+O(\varepsilon^2)\,,
\ee
with $\gamma_{(0,3)}$ given by \eqref{eq:gamma03b} and $\Omega^{KS}$ normalized as in \eqref{eq:OmegaNorm}. Following the same procedure, we obtain the expressions at order $O(\varepsilon)$ for the other components of $G_3$. The $(3,0)$ component is given by:
\be
\label{eq:G30}
G_{(3,0)}=-\frac{8\pi^2\,h^{1/4}}{S^{4/3}\sinh(\tau)^2}\,\pd_\tau(\sinh\tau\,\xi_6-\tau \xi_5)\,\Omega^\KS+O(\varepsilon^2)\,,
\ee
and the $(1,2)$ component by:
\be 
\label{eq:gamma12}
\begin{split}
G_{(1,2)}\,&=\,-\frac{2\sqrt{6}h^{1/4}}{\sinh(\tau)^2}\Bigg[\Big(\xi_5-\text{sech}(\tau)\xi_6 \Big)\Big(h_1\wedge \overline{h}_2\wedge \overline{h}_3 +h_2\wedge \overline{h}_1\wedge \overline{h}_3 \Big.  \\ &+\sinh(\tau)\left(h_1\wedge \overline{h}_1\wedge \overline{h}_3 - h_2\wedge \overline{h}_2 \wedge \overline{h}_3 \right)\Big. \Big) +\left(-\xi_5+ \tau \partial_{\tau}\xi_5 \right. \\
&+\cosh(\tau)\xi_6-\sinh(\tau)\partial_{\tau}\xi_6 \left. \right)h_3\wedge \overline{h}_1\wedge \overline{h}_2 \Bigg. \Bigg] +O(\varepsilon^2)\,.
\end{split}
\ee
Finally, as we have already noted, the $(2,1)$ component is the only one that has a non-vanishing term at zeroth order in $\varepsilon$, given by:
\be 
\label{eq:gamm21}
\begin{split}
G^0_{(2,1)}\,&=\,\frac{2\sqrt{6}}{h^{3/4}}P\text{csch}(\tau)\text{sech} (\tau)\Big(2\coth(\tau)\left(-1+\tau\,\coth(\tau)\right)h_1\wedge h_2 \wedge \overline{h}_3 +\big(\cosh(\tau) \big. \Big. \\
&\!\!\!\!\!\!\!\!\!\!\!\!- \tau\, \text{csch}(\tau)\big)\big(h_1 \wedge h_3 \wedge \overline{h}_1 + \text{csch}(\tau)(h_1\wedge h_3 \wedge \overline{h}_2+h_2\wedge h_3 \wedge \overline{h}_1)-h_2\wedge h_3  \wedge \overline{h}_2  \big. \big) \Big. \Big) \,, 
\end{split}
\ee 
whereas, at first-order in $\varepsilon$, the non-vanishing  $G_{(2,1)}$ components, $G^1_{ij\bar{k}}h_i\wedge h_j \wedge \overline{h}_k$,  are:
\small
\begin{align}
\label{eq:gamma21} 
G^1_{12\bar{3}}&=\frac{\sqrt{3}\,\text{csch}(\tau)}{5\sqrt{2}g_s h^{3/4}}\Big(5\varphi_5+\text{csch}(\tau)\big(4g_sP\left(-1+\tau \coth(\tau)\right)\left(6\varphi_1+4\varphi_3 -5(3\varphi_4 +\varphi_8) \right)\big. \Big. \nonumber \\
&-5(\varphi_5'+\varphi_6')\big.  \big)-5\big(\varphi_6+\coth(\tau)\left(2g_s\varphi_7+\varphi_5'-\varphi_6'\right)-2g_s\varphi_7' \big)\Big. \,,\\
G^1_{13\bar{1}}&=-G_{23\bar{2}}=\frac{\sqrt{6}\text{csch}(\tau)}{5h^{3/4}S^{4/3}}\Big(10h(\xi_5-\xi_6(\text{sech}(\tau)) \Big. \\
&+S^{4/3}\big(P\left(6(1-\tau \text{csch}(\tau)\text{sech}(\tau))\varphi_1 \right. \big. \Big. \nonumber \\
&\left. \big. \Big. +10\tau\, \text{sech}(\tau)\tanh(\tau)\varphi_2-(-1+\tau \text{csch}(\tau)\text{sech}(\tau))(4\varphi_3-15\varphi_4\right)+10\text{sech}(\tau)\varphi_7\big)\Big) \,, \nonumber \\
G^1_{13\bar{2}}&=G_{23\bar{1}}=\frac{2\sqrt{6}\text{csch}(\tau)\text{csch}(2\tau)}{5h^{3/4}S^{4/3}}\Big(10h(\xi_6-\xi_5\cosh(\tau)) \Big. \\
&+S^{4/3}\big(-6P(\cosh(\tau)-\tau \text{csch}(\tau))\varphi_1(\tau) +5P(1+\cosh(2\tau)-4\tau\coth(2\tau))\varphi_2(\tau) \Big.  \nonumber \\
& -P(\cosh(\tau)-\tau\text{csch}(\tau))(4\varphi_3(\tau)-15\varphi_4(\tau))-10\varphi_7(\tau)\big) \Big)\,. \nonumber
\end{align}
\normalsize
Note that the (2,1) flux remains primitive at first order: $G^1_{(2,1)} \wedge J =0 $, where the K\"ahler form is:
\begin{equation}
\label{eq:Kahler form}
J=\frac{1}{2}i\left(h_1\wedge \bar{h}_1+h_2\wedge \bar{h}_2+ h_3\wedge \bar{h}_3 \right)\, .
\end{equation}

The behavior in the UV of all components of $G_3$ can be obtained using the UV expansion of the functions $\varphi_i$ and $\xi_i$ of \cite{Bena:2009xk}. The $(0,3)$ component is given by:  
\be
\label{eq:G03UV}
G_{(0,3)}^{UV}=\left(\frac{2}{3}\right)^{3/4}\,\frac{81\pi\,\varepsilon}{5\,\mathcal{I}_0}\frac{1}{\sqrt{\alpha' M} g_s^{3/4}}\,\log(r_{UV}^3/|S|)^{5/4}\,\left(\frac{|S|}{r_{UV}^3}\right)^{7/3}\,\overline{h}_1\wedge \overline{h}_2\wedge \overline{h}_3\,+\,O\left(\varepsilon^2\right)\,,
\ee
where we introduced a new radial coordinate, $r$:
\be
\label{eq:defr}
r \equiv \frac{3^{1/2}}{2^{5/6}}\,|S|^{1/3}\,e^{\tau/3}\,,
\ee
such that for large values of $\tau$, the six-dimensional metric of the deformed conifold approaches the conifold metric: $dr^2+r^2ds_{T^{11}}^2$.

Note that $G_{(0,3)}$ falls down as $r^{-7}$ in the $\UV$ and is dual to an operator of dimension $\Delta=7$. Such operator is a fermion bilinear of the (schematic) form $F_{\mu \nu}F^{\mu \nu}\lambda\lambda$ where $\lambda$ is the gaugino and $F_{\mu \nu}$ the field strength.

In the same way, we can compute the asymptotic behavior of the $({3,0)}$ component of $G_3$:
\be
\label{eq:G30UV}
G_{(3,0)}^{UV}\,=\,\left(\frac{3}{2}\right)^{29/4}\frac{2\pi\,\varepsilon}{5 \mathcal{I}_0^2}\,\frac{1}{\sqrt{\alpha' M} g_s^{3/4}}\,\log(r_{UV}^3/|S|)^{5/4} \left(\frac{|S|}{r_{UV}^3}\right)^{11/3}\! h_1\wedge h_2\wedge h_3+O(\varepsilon^2)\,.
\ee
The asymptotic decay of $G_{(3,0)}$ indicates that it is dual to an operator of dimension $\Delta=11$ that is again a combination of a fermion bilinear and the field strength of the gauge field of the (schematic) form $(F_{\mu \nu}F^{\mu \nu})^2 \lambda \lambda$. In a similar way, using the asymptotic expansion of the more involved $G_{(1,2)}$ component, we find that the leading term is primitive (as one could also see from \eqref{eq:gamma12}) and is given by: 
\small
\be
\label{eq:G12UV}
G_{(1,2)}^{UV}\!=\!\left(\frac{3}{2}\right)^{7/4}\frac{2\pi \varepsilon}{\mathcal{I}_0}\,\frac{1}{\sqrt{\alpha' M} g_s^{3/4}} \log\!\left(\frac{r_{UV}^3}{|S|}\right)^{1/4} \!\!\!\left(\frac{|S|}{r_{UV}^3}\right)^{4/3}\!\!\!\!\! \left(h_1\wedge \bar{h}_1 \wedge \bar{h}_3 \!-\!h_2 \wedge \bar{h}_2 \wedge \bar{h}_3\right) +O(\varepsilon^2),
\ee
\normalsize
which means that this must be holographically dual to the expectation value of an operator of dimension $\Delta=4$, which corresponds to a marginal deformation of the superpotential \cite{Grana:2001xn}.

Finally, the leading asymptotic terms in the (2,1) component at zeroth order in $\varepsilon$ is given by:
\be
\label{eq:G21UV0}
G_{(2,1)}^{0~UV}=\left( \frac{2}{3}\right)^{1/4}\frac{1}{\sqrt{\alpha' M} g_s^{3/4}}\log \left(\frac{r_{UV}^3}{|S|}\right)^{-3/4}(h_1\wedge h_3\wedge \bar{h}_1-h_2\wedge h_3 \wedge \bar{h}_2)\,,
\ee
whereas the leading asymptotic behavior of the (2,1) component at first order in $\varepsilon$ is 
\be
\label{eq:G21UV1}
G_{(2,1)}^{1~UV}= \frac{24\varepsilon}{\sqrt{\alpha' M} g_s^{3/4}}\log \left(\frac{r_{UV}^3}{|S|}\right)^{-7/4}(h_1\wedge h_3 \wedge \bar{h}_1 - h_2\wedge h_3 \wedge \bar{h}_2)\,. 
\ee

\section{Fermion masses}
\label{sec:gauginomass}

The presence of new components of the fluxes can strongly affect the worldvolume dynamics of D7-branes in the UV. In fact, any component of $G_3$ that is not $(2,1)$ generates fermion masses, possibly breaking supersymmetry on the brane. For instance, a non-trivial $G_{(0,3)}$ component is responsible for a non-vanishing gaugino mass $m_\lambda\neq 0$ \cite{Camara:2004jj,Jockers:2004yj,Grana:2020hyu}. A gaugino mass that is larger than the  gaugino  condensation scale, $\Lambda_{\YM}$, signals a breakdown of the $\mathcal{N}=1$ supersymmetric dynamics at this scale, and affects the expression of the ADS superpotential. Hence, in order for the KKLT {\em ``moduli stabilization via gaugino condensation''} scenario to apply we need {$\frac{m_{\lambda}}{\Lambda_{\YM}} \ll 1$. Our purpose is to calculate this ratio.

The mass of the canonically normalized gaugino is given by \cite{Grana:2020hyu}: 
\be
\label{mlambda1}
m_{\lambda}=\left(-\frac{4\pi^2}{\sqrt{6}}\right)\frac{1}{4}\frac{\int \delta^{(0)}_{\Sigma}\, e^{\phi/2} G_3\wedge \Omega}{\int_{\Sigma}\frac{1}{2}J\wedge J}\,, 
\ee
where we work in Einstein frame and $\delta^{(0)}_{\Sigma}$ localizes the integral on the cycle $\Sigma$ wrapped by the seven-branes. The three-form, $\Omega$, is again normalized such that $\Omega \wedge \bar{\Omega}=\frac{3i}{\pi^4}\text{vol}_6$ and the two-form $J$ such that $\frac{1}{2}J\wedge J=\text{vol}_4$. Here $\text{vol}_6$ is the volume form of the warped Calabi-Yau threefold and $\text{vol}_4$ is the volume of the 4-cycle wrapped by the D7-branes. The $(0,3)$ fluxes \eqref{eq:G03UV} then generate the D7 worldvolume gaugino mass\footnote{Note that this is the fall-off corresponding to an operator of dimension 7 in the holographic theory dual to the KS solution, and not to an operator of dimension 3, which would correspond to a gaugino mass in this theory.}:
\be
\label{mlambda2}
m_{\lambda}= (-i)\left(\frac{2}{3}\right)^{3/4}\!\frac{162\,\pi \,\varepsilon }{5\,\mathcal{I}_0}\frac{g_s^{-1/4}}{\sqrt{\alpha' M} }\log(r_{\UV}^3/|S|)^{5/4}\left(\frac{|S|}{r^3_{\UV}}\right)^{7/3}\, , 
\ee
where $r_{\UV}$ is the radial cut-off, whose value can be determined if we require the warping to be of order one in the UV\footnote{This is the natural assumption if the Calabi-Yau manifold is weakly warped.}.


We now compare this with a four-dimensional supergravity computation of the supersymmetry breaking gaugino masses. In order to compare correctly, we should use the Gukov-Vafa-Witten superpotential, without including the non-perturbative superpotential coming from gaugino condensation on the D7-branes, since the computation of the $\aDt$ brane induced fluxes is performed without it.   
The general form of Majorana gaugino masses in four-dimensional supergravity is given, up to an irrelevant phase, by
\begin{equation}
m_{\lambda} = \frac{1}{2 Re f} {\cal K}^{i \bar\jmath} e^{\frac{{\cal K}}{2}} \overline{D_{\bar\jmath} W} \frac{\partial f}{\partial z^i} \ , \label{4d1}
\end{equation}
where ${\cal K}$ is the K\"ahler potential, $D_i W=\partial_i W+(\partial_i {\cal K}) W$ is the K\"ahler covariant derivative of the superpotential $W$, $f (z^i)$ is the gauge kinetic function and $z^i$ denote the complex scalars in the chiral multiplets.
 
The relevant quantities in the 4d SUGRA Lagrangian for a single K\"ahler modulus are\footnote{For more details of this particular form of the K\"ahler potential and superpotential see \cite{Bena:2009xk}. }
\begin{eqnarray}
{\cal K} = - 3 \log r -\log(2/g_s) - \log\left(\frac{|\Omega|^2 V_w^2}{\kappa_4^{12}} \right) \,, \hspace{3pt} W= \frac{1}{\kappa_4^8}\int_M G_3 \wedge \Omega \,,  \hspace{3pt}  f_{D7} = T \,, \label{4d2} 
\end{eqnarray}
where
\begin{equation}
 r  =  T + {\bar T} - \frac{3c' g_s (\alpha' M)^2}{\pi |\Omega|^2 V_w } |S|^{2/3} \, .  \label{4d3} 
\end{equation}
Here $V_w$ is a fiducial volume that we take to be equal to one, $\kappa_4=M_{pl}^{-2}$, $c'=1.18$ is a numerical factor coming from taking into account warping effects in the effective field theory \cite{Douglas:2007tu} and $\Omega$ has the usual normalization: $|\Omega|^2=8$. From \eqref{4d2} one can easily see that
\begin{equation}
\label{4d3a}
D_T W = - \frac{3}{\kappa_4^8 r} \int G_3 \wedge \Omega \,.
\end{equation}
Combining all the above relations, the gaugino mass \eqref{4d1} is 
\begin{equation}
m_{\lambda} = - \frac{\sqrt{g_s}}{ 4\kappa_4^8 (T + {\bar T}) r^{1/2}} \int G_3 \wedge \Omega  \,. \label{4d4}
\end{equation}
In order to compare this expression with \eqref{mlambda2} we should note that in Einstein frame, $T=\frac{\text{vol}_4}{2\pi \alpha'^2}$ and that for a single K\"ahler modulus $\text{vol}_6=\frac{\sqrt{2}}{3}T^{3/2}$. Taking into account the $ \delta^{(0)}_{\Sigma}$ localization factor in \eqref{mlambda1}, it is not hard to see that, in the regime where we can neglect the conifold  contribution in \eqref{4d3}, the two results agree parametrically.\footnote{An independent check of the agreement between the 10d versus the 4d description would be a ten-dimensional computation of the gravitino mass, along the lines of \cite{DeWolfe:2002nn}, taking into account the backreaction effects from the antibrane. }

\subsection{Gaugino mass and gaugino condensation}
As already mentioned, the component $G_{(0,3)}$ is non-vanishing and gives rise to a mass for the world-volume gaugino of the $\Dse$-branes. Since this mass is generated by the addition of $\aDt$ branes in the IR throat, we can think of it as the manifestation of a $\aDt$-$\Dse$ interaction. In order to see whether and how this mass can affect the dynamics of the gaugino condensation responsible for K\"ahler-moduli stabilization in the KKLT scenario, we need to compute the various energy scales of our system. In the following, we will mostly follow the same conventions as in \cite{Bena:2018fqc,Bena:2019sxm}. First let us recall that:
\be
M\,=\,\frac{1}{4\pi^2\alpha'}\,\int_A\,F_3\,,\quad  K\,=\,\frac{1}{4\pi^2\alpha'}\,\int_B\,H_3\,,
\ee
where the $A$-cycle is the $\,\mathrm{S}^3$ at the bottom of the throat. The $B$-cycle is a bit more subtle, extending to the brim of the KS throat and into the Calabi-Yau compactification manifold. Here we approximate $K$ with the integral of $H_3$ over the part of this cycle inside the throat, ignoring the contribution from the rest of the compactification manifold. 

If we call $\tau_{UV}$ the distance from the tip of the KS throat to the region where the throat	 merges with the CY\footnote{In KS holography this corresponds to the UV cutoff, $\Lambda_{UV}$ .}, then
\be
K\,=\,\frac{1}{4\pi^2\alpha'}\int_{0}^{\tau_{UV}}\!\!\!\!\mathrm{d}\tau\!\int_{\mathrm{S}^2}H_3\,,
\ee
where the two-sphere is taken at $\psi=0,\,\theta_1=-\theta_2,\,\phi_1=-\phi_2$. Using the form \eqref{eq:fluxesKS} for the fluxes and working at zeroth order in $\varepsilon$, we can compute the units of NSNS-flux as:
\be
K=-\frac{2}{\pi\alpha'} f(\tau)\bigg{|}^{\tau_{UV}}_{0}\approx \frac{g_s M}{2\pi}\,\tau_{UV}\,,
\ee
where in the last expression we evaluated $f(\tau)$ assuming $\tau_{UV}$ to be large. This gives the radial cutoff of the throat and its associated energy scale :
\be
\label{eq:deftauUV}
r^3_{UV}=\Lambda^3_{UV}\,=\,\frac{3^{3/2}}{2^{5/2}}\,|S|\,e^{\frac{2\pi K}{g_s M}}\,,
\ee
where we used the relation between $\tau$ and $r$ given in \eqref{eq:defr}. Using holographic KS terminology, one can also define an infrared energy scale, which can be taken to coincide with the value of $r$ at the bottom of the throat ($\tau=0$). We can thus introduce a parameter, $\eta$, measuring the hierarchy between the ultraviolet and infrared scales:
\be
\eta\,=\,3\ln \frac{\Lambda_{UV}}{\Lambda_{IR}}\,=\,\frac{2\pi K}{g_sM}\,.
\ee

Even if we do not know the full metric and fluxes of the Calabi-Yau compactification, it is reasonable to assume that, when the size of the compactification manifold is not very large, the fields at the location of the $N_{\Dse}$ seven-branes wrapping a four-cycle $\Sigma$ are of the same order as the fields at the brim of the KS throat, at $r_{UV}=\Lambda_{UV}$. 

In the absence of a gaugino mass, the low-energy world-volume theory of $\Dse$ branes at a non-singular locus is 8-dimensional $SU(N_{\Dse})$ SYM theory, which one further compactifies on $\Sigma$\footnote{When the D7 branes are on top of an $\mathrm{O}7$-plane, the gauge group can be orthogonal or symplectic.}. The coupling constant of the resulting ${\cal N}=1$ four-dimensional gauge theory runs logarithmically with the energy and the theory confines in the infrared at a scale
\be
\label{eq:lambdaYM}
\Lambda_{\YM}\,\approx\,\mu_0 e^{-\frac{2\pi \Re T}{3N_{D7}}} \,,
\ee
where $\mu_0$ is the ``UV scale'' that depends on (stabilized) complex-structure moduli and will be assumed to be of order one (in Planck units). If the world-volume theory has a more general gauge group, $G$, the exponent in \eqref{eq:lambdaYM} must be replaced by $-2\pi \Re T/(3\,\#_C(G))$, with $\#_C(G)$ the dual Coxeter number of $G$. Furthermore, in the confined phase the gaugino condenses, giving rise to a nontrivial contribution to the superpotential that depends on the coupling constant of the four-dimensional ${\cal N}=1$ theory, and hence on the volume of the four-cycle wrapped by the D7 branes. 

In the presence of gaugino mass, $m_{\lambda}$, induced by the three-form fields sourced by the antibranes, this scenario can change. In particular, when $m_{\lambda}$ is larger than $\Lambda_{\YM}$, the theory will confine at an energy scale proportional $m_{\lambda}$, and the resulting potential will be independent of the four-cycle volume, thus ruining the K\"ahler moduli stabilization. Hence, gaugino condensation can only stabilize the K\"ahler moduli when:
\be
\label{eq:scaleProblem}
\frac{m_{\lambda}}{\Lambda_{\text{YM}}} \ll 1\,.
\ee

To estimate this ratio one can consider a KKLT-like scenario, in which the value of the potential at the AdS minimum is proportional to the square of the non-perturbative superpotential, while the contribution to the scalar potential from the anti-D3 branes (which sit at the tip of the throat, where their energy is minimized) is proportional to $H^{-1}$. In a realistic de Sitter compactification, where we want the cosmological constant to be as small as possible, these two quantities should have approximately the same magnitude, which means that the following relation should hold\footnote{ If one takes into account the explicit expression for both $V_{\overline{D3} }$ \cite{Dudas:2019pls} and $V_{AdS}$, one finds using \eqref{eq:SwithQ} a more precise expression
\be
 e^{- \frac{2\pi \Re T}{3 N_c}}  \approx  \left( \frac{N_c}{A } \right)^{1/3} \left( \frac{ \eta}{\Re T}\right)^{1/6}  e^{-2 \eta /9} \,, \label{footnote1}
 \ee where $A$ is the Pfaffian factor that appears in the non-perturbative superpotential. Since we assume both $A$ and $N_c$ to be of order one, the difference between the results obtained with \eqref{footnote1} and those obtained with \eqref{eq:aD3=NP} is negligible.}  
\be
\label{eq:aD3=NP}
V_{\overline{D3}}\,\approx\, V_{\text{AdS}} \rightarrow  \frac{|S|^{4/3}}{h_0}\,\approx \, e^{-\frac{4\pi}{N_c}\Re(T)}\,.
\ee

In our conventions the hierarchy is given by $e^{-\eta}$ and we note that  $\tau_{\UV}\approx \eta$. We also require that the warping is $O(1)$ in the $UV$. All these give a relation between the complex-structure modulus, $S$, and the flux-induced D3 charge of the throat, $Q_{\Dt}^{\text{Throat}}=K M$:
\be
\label{eq:SwithQ}
H\,=\,\frac{h(\tau)}{|S|^{4/3}}\approx 1\quad \rightarrow \quad |S|^{4/3}\approx h(\tau_{UV})\,\approx\,6\,\pi \cdot 2^{1/3} (\alpha')^2 e^{-\frac{4\eta}{3}}\,Q_{\Dt}^{\text{Throat}}\,.
\ee
Combining \eqref{eq:aD3=NP} (more precisely \eqref{footnote1}) and \eqref{eq:SwithQ}, we can estimate the value of the stabilized K\"ahler modulus 
\be
\label{eq:Kahler modulus}
\Re T\sim \frac{2 N_c K}{3 g_sM}\,=\,\frac{N_c}{3\pi}\eta\,,
\ee
where we have ignored logarithmic corrections coming from the non-exponential terms in \eqref{footnote1}.

Using this approximation, we can evaluate the ratio between the gaugino mass and the condensation scale in the KKLT scenario:
 \be
\frac{m_\lambda}{\Lambda_{\YM}} \approx 10^2\overline{N} M^{-5/2}g_s^{-5/4}\eta^{5/4}e^{-\frac{19}{9}\eta} \,.
 \ee
Our calculation indicates that this ratio must be smaller than one in all flux compactifications where K\"ahler moduli are stabilized via D7 gaugino condensation and where the cosmological constant is uplifted using anti-D3 branes in warped Klebanov-Strassler-like throats. It is not hard to see that for the range of parameters one uses in ``vanilla'' KKLT scenarios this bound will be satisfied. For example for the de Sitter minimum of \citep{Blumenhagen:2019qcg}, where $M=K=70$ and $g_s=1/2$ we find this ratio to be: 
\be
\frac{m_\lambda}{\Lambda_{\YM}} \sim 10^{-13} \,.
\ee
Hence, in these scenarios, the stabilization of the K\"ahler moduli via D7 gaugino condensation is not affected by the fluxes produced by the anti-D3 branes. 

It would be interesting to try to span larger families of possible flux compactifications and other de Sitter scenarios which use a smaller hierarchy (such as those of \cite{Bento:2021nbb,Crino:2020qwk}) to see whether this bound may become harder to satisfy and may result in nontrivial constraints on the parameters of the compactification.


\acknowledgments We would like to thank Veronica Collazuol, Nicol\'as Kovensky, Severin L\"ust and Salvatore Raucci for useful discussions.
This work was supported in part by the ERC Grants 772408 ``Stringlandscape'' and 787320 ``QBH Structure'', and by the John Templeton Foundation grant 61149.


\bibliographystyle{ytphys}
\bibliography{ref}

\end{document}